\documentclass[floats,floatfix,showpacs,amssymb,prd,superscriptaddress,twocolumn,aps]{revtex4-1}

\usepackage{amssymb,amsmath}
\usepackage{amsthm}
\usepackage{graphicx}
\usepackage{hyperref}
\usepackage{epstopdf}

\begin{document}

\title{Scalar and axial quasinormal modes of massive static phantom wormholes}
\author{Jose Luis Bl\'azquez-Salcedo}
\email[{\it Email:}]{jose.blazquez.salcedo@uni-oldenburg.de}
\author{Xiao Yan Chew}
\email[{\it Email:}]{xiao.yan.chew@uni-oldenburg.de}
\author{Jutta Kunz}
\email[{\it Email:}]{jutta.kunz@uni-oldenburg.de}
\affiliation{
Institut f\"ur  Physik, Universit\"at Oldenburg, Postfach 2503,
  D-26111 Oldenburg, Germany}

\date{\today}

\begin{abstract}
We study the quasinormal modes for scalar, axial and radial perturbations
of massive static spherically symmetric wormholes supported by a phantom
scalar field, calculating the modes directly and employing the WKB
approximation.
The spectrum of the quasinormal modes is compared with the 
spectrum of Schwarzschild black holes. For fixed multipole number $l$ and 
large mass $M$ the wormhole modes approach their black hole counterparts
as $1/M$. 
\end{abstract}

\maketitle





\section{Introduction}

Wormholes represent hypothetical astrophysical objects connecting 
two distant regions within the Universe (intra-universe wormholes)
or two distinct asymptotic regions (inter-universe wormholes)
\cite{Morris:1988cz,Visser:1995cc,Lobo:2017eum}.
In the simplest case wormholes feature a single throat,
which constitutes a minimal surface of the spacetime.

The probably best known example of a non-traversable wormhole 
is the Einstein--Rosen bridge, which actually is a special case
of the Schwarzschild black hole,
obtained %
by performing a coordinate transformation \cite{Einstein:1935tc}. 
In contrast, the construction of traversable wormholes 
in general relativity (GR) 
requires the presence of exotic matter \cite{Morris:1988cz}
(at least in the case of minimally coupled fields).
Let us note, that phantom matter can be applied in cosmology to
describe the accelerated expansion of the Universe
(see e.g.~\cite{Lobo:2005us}).

The static spherically symmetric Ellis wormholes
\cite{Ellis:1973yv,Bronnikov:1973fh,Ellis:1979bh}
are the simplest examples 
of such traversable wormholes which are supported by a phantom field,
i.e., a form of exotic matter.
Ellis wormholes have been generalized to higher dimensions
\cite{Torii:2013xba,Dzhunushaliev:2013jja}
and they have been put into rotation
\cite{Kashargin:2007mm,Kashargin:2008pk,Kleihaus:2014dla,Chew:2016epf},
they have been embedded in scalar--tensor theory \cite{Chew:2018vjp},
and they have been generalized to feature a complex phantom field
\cite{Dzhunushaliev:2017syc}.

A number of observational signatures have been proposed for astrophysical
searches for wormholes, which include their gravitational lensing
\cite{Cramer:1994qj,Safonova:2001vz,Perlick:2003vg,Nandi:2006ds,Abe:2010ap,Toki:2011zu,Nakajima:2012pu,Tsukamoto:2012xs,Kuhfittig:2013hva,Bambi:2013nla,Takahashi:2013jqa,Tsukamoto:2016zdu},
their shadows
\cite{Bambi:2013nla,Nedkova:2013msa,Ohgami:2015nra,Shaikh:2018kfv,Gyulchev:2018fmd},
signatures of their accretion disks
\cite{Harko:2008vy,Harko:2009xf,Bambi:2013jda,Zhou:2016koy,Lamy:2018zvj},
or their possible distinguishing features as black hole alternatives
(see e.g. \cite{Damour:2007ap,Bambi:2013nla,Azreg-Ainou:2014dwa,Dzhunushaliev:2016ylj,Cardoso:2016rao,Konoplya:2016hmd,Nandi:2016uzg,Bueno:2017hyj}).
Moreover, there are recent studies on new configurations
involving wormholes, such as neutron star--wormhole systems
\cite{Dzhunushaliev:2011xx,Dzhunushaliev:2012ke,Dzhunushaliev:2013lna,Dzhunushaliev:2014mza,Aringazin:2014rva},
or boson star--wormhole systems
\cite{Dzhunushaliev:2014bya,Hoffmann:2017jfs,Hoffmann:2017vkf,Hoffmann:2018}.

The recent detection of gravitational waves by LIGO/Virgo from the merger 
of binary black holes 
\cite{Abbott:2016blz,Abbott:2016nmj,Abbott:2017vtc,Abbott:2017oio}, 
and binary neutron stars 
\cite{TheLIGOScientific:2017qsa} 
has made the investigation of the emission of gravitational waves
from massive compact objects highly relevant.
In this connection wormholes have received
much attention, since they might mimick black holes
\cite{Damour:2007ap,Bambi:2013nla,Azreg-Ainou:2014dwa,Dzhunushaliev:2016ylj,Cardoso:2016rao,Konoplya:2016hmd,Nandi:2016uzg,Bueno:2017hyj}.
In particular, 
quasinormal modes of several types of wormholes have been considered
to some extent
\cite{Kim:2008zzj,Konoplya:2005et,Konoplya:2010kv,Konoplya:2016hmd,Bueno:2017hyj,Aneesh:2018hlp,Volkel:2018hwb}. 

A quasinormal mode analysis not only yields the ringdown modes 
of a wormhole, it also reveals the 
stability or instability of the perturbed object.
In the case of the simplest traversable wormholes, the Ellis wormholes,
it has been known for some time, that they possess an unstable 
radial mode
\cite{Shinkai:2002gv,Gonzalez:2008wd,Gonzalez:2008xk,Cremona:2018wkj}.
But a quasinormal mode analysis has so far only been performed
(in part) for the symmetric, massless Ellis wormhole \cite{Kim:2008zzj}.
A study of the physically more interesting massive case
has not yet been done.
However, since astrophysical objects are expected to rotate,
the true goal should be to investigate the quasinormal modes
of rotating Ellis wormholes, which might even be free of the
radial instability of the static wormholes
\cite{Matos:2005uh,Dzhunushaliev:2013jja}.

As a very first step towards this goal,
in the present paper we calculate the quasinormal modes 
of massive static phantom wormholes numerically by direct integration and with the WKB method.
We show that in the massless case
our numerical results are in close agreement with Kim's results,
who calculated the fundamental scalar ($l>0$), electromagnetic 
and axial ($l>2$) quasinormal modes 
employing the WKB method \cite{Kim:2008zzj}.
We also compare the quasinormal mode spectrum of the Ellis wormholes
with the one of the Schwarzschild black holes.  

\section{Phantom Wormholes}

We consider GR coupled minimally to a massless phantom field
$\phi$. The action is then given by
\begin{equation}
  S=\frac{1}{16 \pi G} \int d^4 x \sqrt{-g} 
\left[ R + 2 \nabla_\mu \phi \nabla^\mu \phi   \right]\,.
\end{equation}
By varying the action with respect to the metric and the phantom field, 
the equation of motions are obtained
\begin{align}
R_{\mu \nu} = - 2 \partial_\mu \phi \partial_\nu \phi \,,  \\ 
 \nabla_\mu \nabla^\mu \phi  = 0 \,. \label{KGeqn}
\end{align}

The solutions for the static spherically symmetric Ellis wormholes 
are known in closed form \cite{Ellis:1973yv,Bronnikov:1973fh,Ellis:1979bh}
\begin{align}
   \phi   &= \frac{Q}{r_0} \left[ \tan^{-1} \left( \frac{r}{r_0} \right) -\frac{\pi}{2}  \right] \,, \\
 ds^2 &= -e^{f} dt^2 + \frac{1}{e^f} \left[ dr^2 + (r^2+r_0^2) (d\theta^2 + \sin^2 \theta d\varphi^2) \right] \,,
\end{align}
with
\begin{equation}
 f = \frac{C}{r_0} \left[ \tan^{-1} \left( \frac{r}{r_0} \right) -\frac{\pi}{2}  \right] \,,
\end{equation}
where $r_0$, $Q$ and $C$ are constants. The field equations impose $4 Q^2 = C^2 + 4r_0^2$.
The radial coordinate $r$ ranges from $-\infty$ to $\infty$.

As $r \rightarrow +\infty$, $f \rightarrow 0$, 
and the metric evidently approaches Minkowski spacetime.  
As $r \rightarrow -\infty$,
a coordinate transformation 
 \begin{equation} \label{coor_trans}
  \bar{t} = e^{-\frac{C \pi}{2 r_0}} t \,, \quad \bar{r} = e^{\frac{C \pi}{2 r_0}} r \,, \quad \bar{r}_0 = e^{\frac{C \pi}{2 r_0}} r_0  \,,
 \end{equation}
reveals asymptotic flatness in this limit as well.
Thus the spacetime has two asymptotically flat regions, which are
connected by a throat, as an inspection of the 
circumferential radius $R(r)$,
\begin{equation}
R^2(r) = e^{-f}(r^2 + r_0^2) \,,
\end{equation}
shows.

Concerning the global charges,
the constant $Q$ denotes the charge of the phantom field. 
The constant $C$ can be read off from the asymptotic expansion 
at $r \rightarrow  \pm \infty$ as
\begin{equation}
 C =  2 M \,, 
\end{equation}
where $M$ denotes the mass extracted in the asymptotically flat region,
$r \to +\infty$. 

\begin{figure}
\centering
\includegraphics[width=0.15\textwidth,trim = 70 50 70 50]{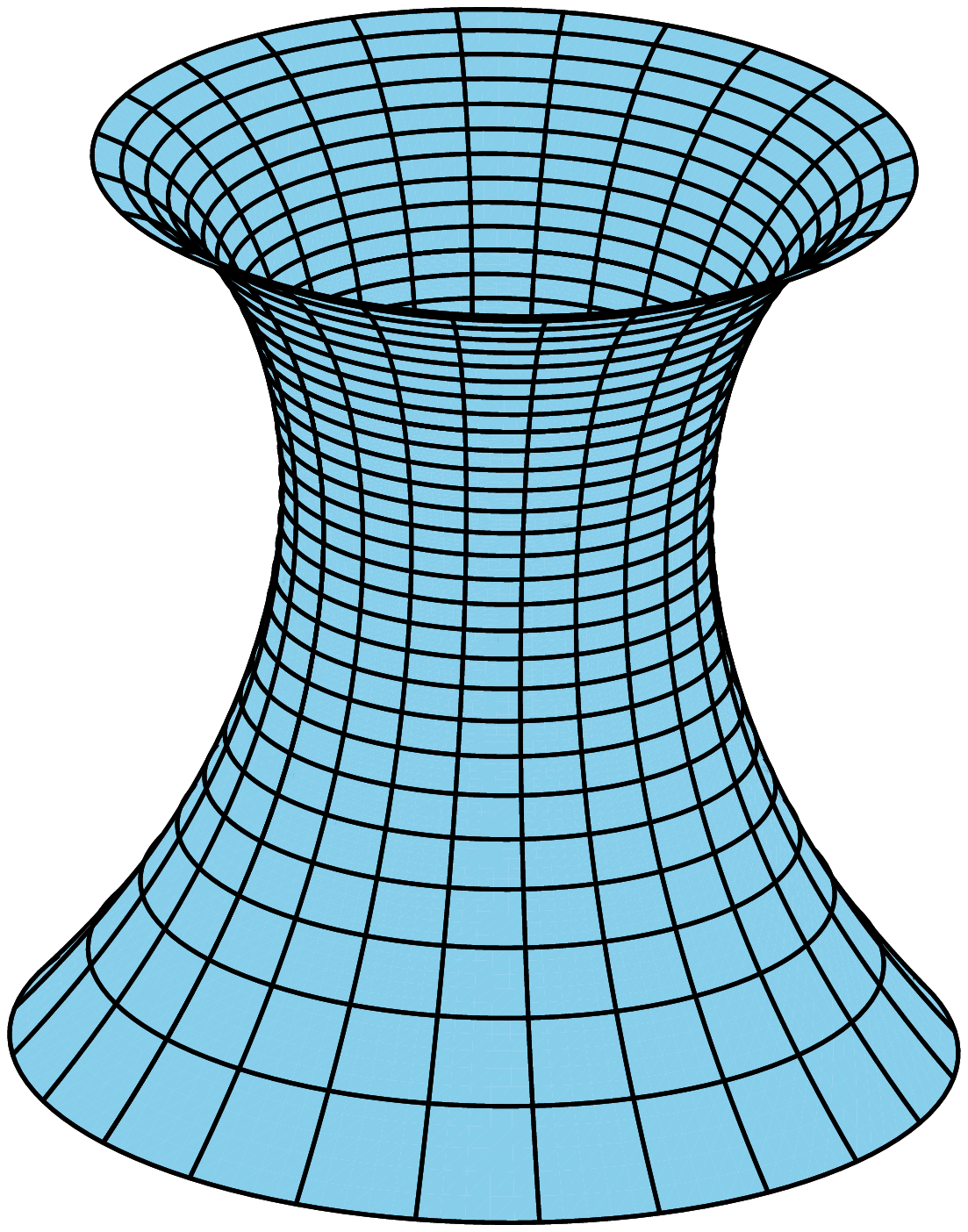}
\includegraphics[width=0.15\textwidth,trim = 70 50 63 50]{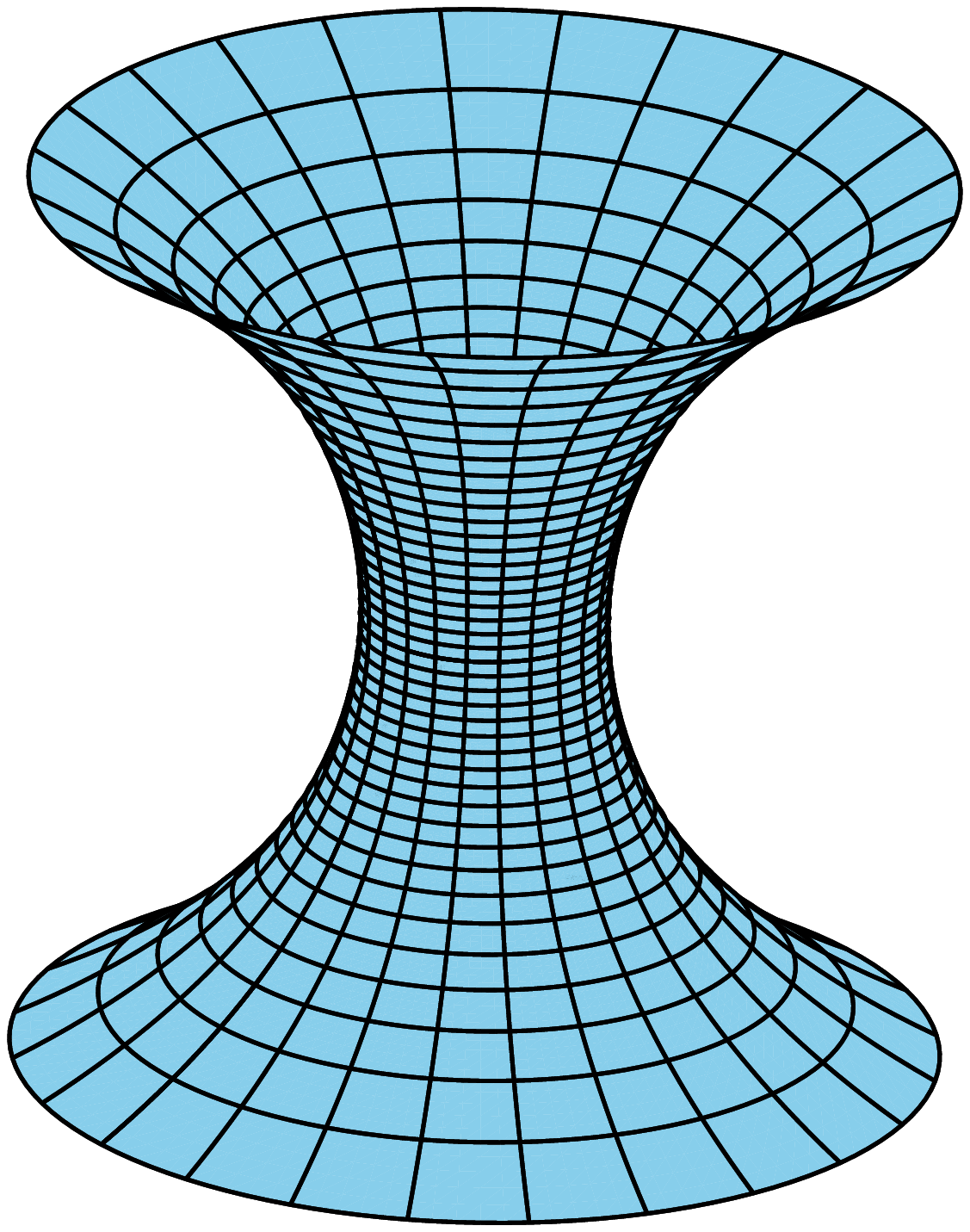}
\includegraphics[width=0.15\textwidth,trim = 70 50 70 50]{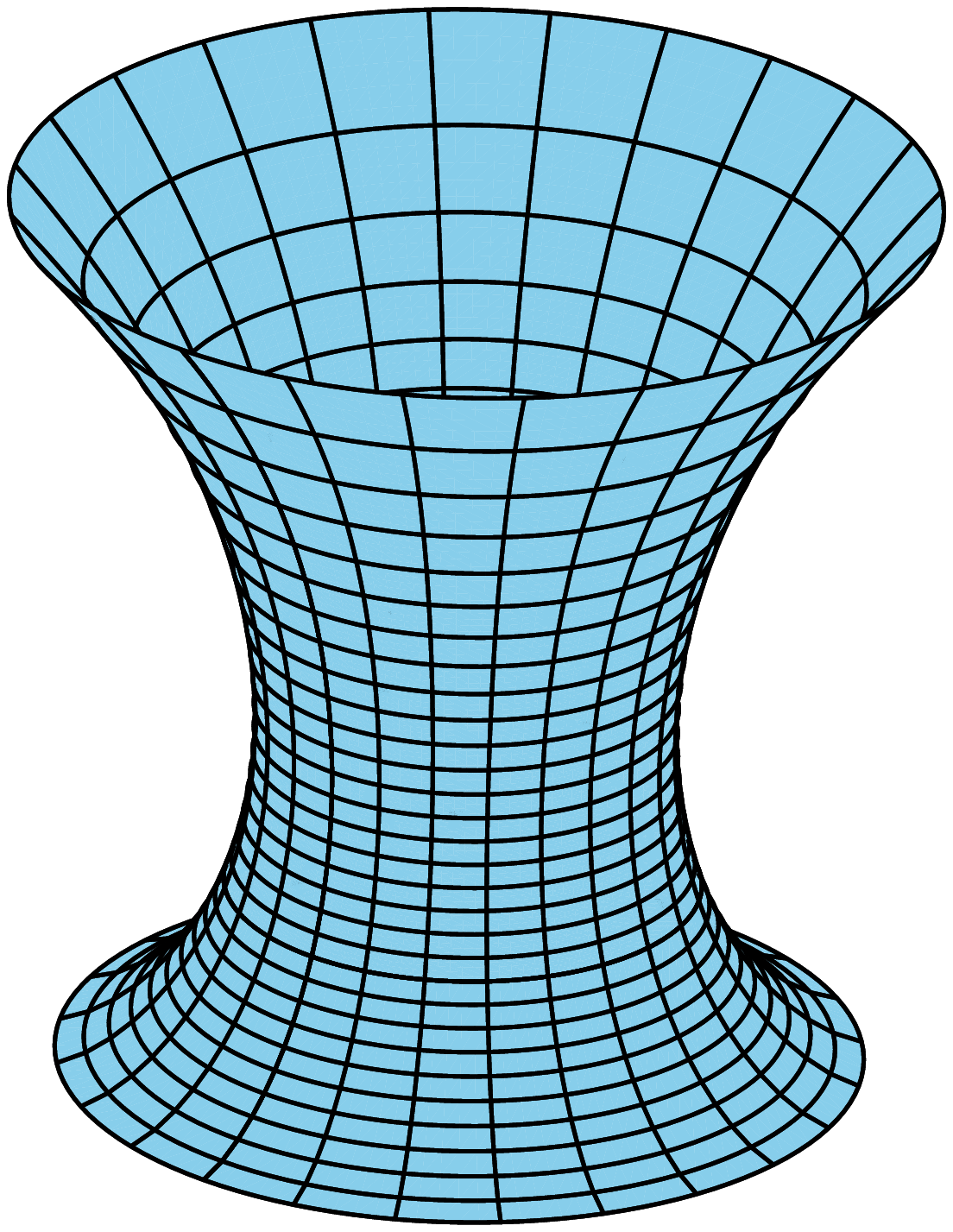}
\caption{Embedding diagrams of wormholes with $C=-1,0,1$, respectively,
from left to right.
}
\label{Fig:embedding}
\end{figure}

The constant $C$ represents also a measure of the symmetry of wormhole. 
The wormhole is symmetric with respect to $r=0$ when $C=0$,
and the wormhole is asymmetric when $C \neq 0$.
The symmetry/asymmetry is visualized 
in the embedding diagrams in Fig.~\ref{Fig:embedding}.

It is interesting to note that there exists a relation between solutions with a positive value of $C$ and with a negative value,
\begin{eqnarray}
f(r,C) = f(-r,-C) - \frac{\pi C}{r_0}, \nonumber \\
\phi(r,C) = -\phi(-r,-C) - \frac{\pi Q}{r_0}.
\end{eqnarray}
We will see that this relation relates the part of the spectrum of 
quasinormal modes with $C<0$ with the $C>0$ part.


\section{Perturbations of the wormhole}

In this section we will briefly introduce the equations describing perturbations of three different types. 
First we will consider scalar perturbations, 
where only the scalar field is perturbed with no backreaction from the metric. 
Then we will consider axial perturbations, 
which do not couple to the phantom field. 
Finally we will address radial perturbations, 
which couple to both the metric and the phantom field, 
and give rise to the well-known radial instability
\cite{Shinkai:2002gv,Gonzalez:2008wd,Gonzalez:2008xk}.

\subsection{Scalar Perturbations}

The perturbation of the phantom field over the static background WH can be written like
\begin{equation}
\phi = \phi^{(0)} + \epsilon \delta \phi \,.
\end{equation}
The ansatz for the scalar perturbations is
\begin{equation}
  \delta \phi = u(r) e^{-i \omega t} P_l (\cos \theta) e^{i m \varphi} \,,
  \label{pert_phi}
\end{equation}
with $P_l (\cos \theta)$ a Legendre polynomial.

The equation of the perturbation is determined completely by the Klein--Gordon equation (Eq. \eqref{KGeqn}), and can be written as a second--order ODE
\begin{equation} \label{ODEKG}
 \frac{d^2 u}{d r^2} + \frac{2 r}{r^2 + r_0^2} \frac{d u}{d r} + \left[  \frac{\omega^2}{e^{2 f}} - \frac{l (l+1)}{r^2 + r_0^2} \right] u = 0\,.
\end{equation}

Let $u(r)=Z(r) e^{f/2}(r^2+r_0^2)^{-1/2}$, then the above ODE can be written as
a Schr\"{o}dinger--like equation,
\begin{equation}
 \frac{d^2 Z}{d r_{*}^2} + (\omega^2 - V_s(r) ) Z =0 \,,
 \label{Z_scalar}
\end{equation}
with effective potential $V_s(r)$,
 \begin{eqnarray}
  V_s(r) = e^{2 f} \left(   \frac{l(l+1)+1}{r^2+r_0^2} - \frac{(2r-C)^2}{4(r^2+r_0^2)^2} \right)  \,,
\label{V_s}
\end{eqnarray}
and tortoise coordinate $r_{*}$,
 \begin{equation}
 \frac{d r_{*}}{d r}  = e^{-f} \,,
 \label{tortoise}
\end{equation}
where the tortoise coordinate is chosen to have the same sign as the radial coordinate $r$.

\begin{figure}
\centering
\includegraphics[angle =-90,width=0.48\textwidth]{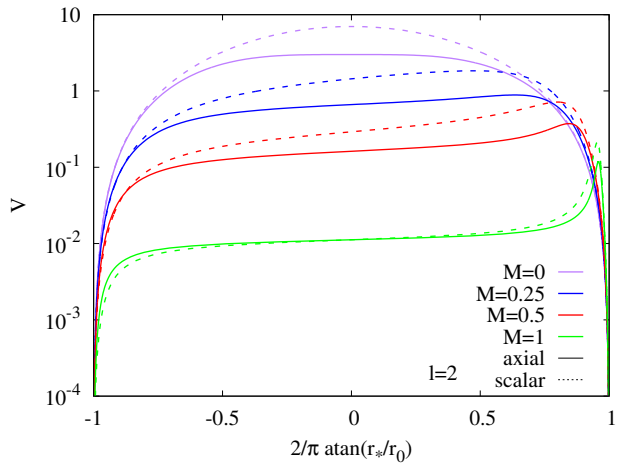}
\caption{Effective potential $V$ vs the compactified tortoise coordinate 
$2/\pi \arctan{(r_*/r_0)}$ for $l=2$, $r_0=1$ 
and several values of the mass $M$.
Dashed lines correspond to the scalar effective potential $V_s$, 
solid lines to the axial one $V_A$.}
\label{Fig:potential}
\end{figure}

In Fig.~\ref{Fig:potential} we show typical profiles of the scalar potential 
$V_s$ (dashed lines)
vs the compactified tortoise coordinate $2/\pi \arctan{(r_*/r_0)}$
for multipole number $l=2$, $r_0=1$ and several values of the mass $M$.

For a fixed value of $l$ and sufficiently large values of the parameter $C$, 
the potential becomes negative at some point. This happens when
\begin{equation}
|C| > 2 r_0 \sqrt{l(l+1)}.
\end{equation}
However the integral 
\begin{eqnarray}
 \int^{\infty}_{-\infty} V_s(R) dR = \frac{1-e^{\pi C/r_0}}{C}[l(l+1)+1/2] 
\end{eqnarray}
does not become negative, and we will see that there is no sign of unstable modes for these perturbations.

Note that with respect to the sign of $C$, the relations 
 \begin{eqnarray}
 dr_{*}(r,C)  = e^{\pi C/r_0} dr_{*}(-r,-C)\,, \nonumber \\
 V_s(r,C) = e^{2 \pi C/r_0} V_s(-r,-C)\,, 
 \label{rel_negC_V}
\end{eqnarray}
imply 
\begin{eqnarray}
\omega(C) = e^{\pi C/r_0} \omega(-C)\,.
\label{omega_sym}
\end{eqnarray}

\subsection{Axial Perturbations}

For a small perturbation $h_{\mu \nu}$ 
in the background spacetime $g_{\mu \nu}^{(0)}$, 
the full metric can be written as
\begin{equation}
 g_{\mu \nu} = g_{\mu \nu}^{(0)} + \epsilon h_{\mu \nu} \,,
\end{equation}
 where $\epsilon$ is an infinitesimal positive number.

The odd--parity perturbation $h_{\mu \nu}$ then 
takes the form \cite{Cardoso:2001bb}
\begin{equation}
h_{\mu \nu}=
\begin{pmatrix}
0   &    0   &    0  &  h_0(r) S_{\theta}    \\
0   &    0   &    0   &  h_1(r) S_{\theta}    \\
0   &    0   &    0    &   0          \\
h_0(r) S_{\theta}  &  h_1(r) S_{\theta} &  0  &  0          
\end{pmatrix} e^{-i \omega t} 
\end{equation}
with $S_{\theta}=e^{i m \varphi}\sin \theta \partial_{\theta}P_l(\cos{\theta})$.

The perturbation is determined by the Einstein equations, 
which reduce to the following minimal set of ODEs
\begin{align}
  \frac{d h_0(r)}{d r} &= \frac{2 r-C}{r^2+r_0^2}  h_0(r)  -i \left( e^{2 f} \frac{ l (l+1) -2}{\omega (r^2+r_0^2)} - \omega \right)  h_1 (r)  \,,  \nonumber \\
 \frac{d h_1(r)}{d r} &= - i \omega e^{-2f} h_0 (r) - \frac{C}{r^2+r_0^2}  h_1(r)  \,.
\end{align}

Defining $h_1(r)=Z(r) e^{-3f/2}(r^2+r_0^2)^{1/2}$, 
this system of equations leads to a Schr\"{o}dinger--like equation,
\begin{equation}
 \frac{d^2 Z}{d r_{*}^2} + (\omega^2 - V_A(r) ) Z =0 \,,
 \label{Z_axial}
\end{equation}
with effective potential $V_A$,
\begin{eqnarray}
  V_A(r) = e^{2 f} \left(   \frac{l(l+1)-3}{r^2+r_0^2} + \frac{3(2r-C)^2}{4(r^2+r_0^2)^2} \right)  \,,
 \label{V_A}
\end{eqnarray}
and the tortoise coordinate Eq.~(\ref{tortoise}).

In Fig.~\ref{Fig:potential} we show typical profiles for 
the axial potential $V_A$ (solid lines). 
Note that the axial and the scalar potentials show the same basic behaviour.
Therefore one can expect the qualitative properties 
of both quasinormal mode spectra to be similar.

Similar to the scalar case the integral 
\begin{eqnarray}
 \int^{\infty}_{-\infty} V_a(R) dR = \frac{1-e^{\pi C/r_0}}{C}[l(l+1)-3/2] 
\end{eqnarray}
is always positive, and we don't find any trace of unstable modes for these perturbations.
It is straightforward to check that the eigenvalue $\omega$ 
satisfies also the symmetry relation (\ref{omega_sym}).

\subsection{Radial Perturbations}

In the case of radial perturbations, we perturb both the metric and the phantom field
\begin{eqnarray}
 g_{\mu \nu} = g_{\mu \nu}^{(0)} + \epsilon h_{\mu \nu} \,, \nonumber \\
 \Phi = \phi(r) + \epsilon \delta \phi (r) e^{i \omega t} \,,
\end{eqnarray}
where now the metric perturbation is $h_{\mu \nu}=diag(e^{f}F_0(r),e^{-f}F_1(r),e^{-f}F_2(r)r^2,e^{-f}F_2(r)r^2 \sin{\theta}^2)$ and the scalar perturbation is as in Eq.~(\ref{pert_phi}) with $l=0$. 

The gauge freedom is fixed by the condition $2 F_2+F_0-F_1 = \delta \phi = 0$. Then the perturbation equations are given by
\begin{eqnarray}
&&\frac{d F_0}{d r} = \left[ \frac{2(r^2+r_0^2)e^{-2f}}{2 r-C}\omega^2 + \frac{2r_0^2}{(r^2+r_0^2)(C-2r)} \right]  F_2 \nonumber \\ &&+ \frac{Cr+2r_0^2}{(r^2+r_0^2)(2 r-C)}F_0   \,, \nonumber \\
&&\frac{d F_2}{d r} =  \frac{1}{2}\frac{C-2r}{r^2+r_0^2} F_0 + \frac{r}{r^2+r_0^2} F_2 \,. 
\label{eq_rad}
\end{eqnarray}

Defining $F_2(r) = Z(r) \frac{C-2r}{\sqrt{r^2+r_0^2}}e^{C\pi/(4r_0)}e^{f/2}$, 
we obtain
\begin{equation}
 \frac{d^2 Z}{d r_{*}^2} + (\omega^2 - V_R(r) ) Z =0 \,,
 \label{Z_radial}
\end{equation}
with the effective potential $V_R$,
\begin{equation}
    V_R(r) = \frac{e^{2 f}}{(r-C/2)^2}\left[ 2 - \frac{r^2+Cr+3r_0^2}{r^2+r_0^2} - \frac{(r-C/2)^4}{(r^2+r_0^2)^2} \right] \,.
 \label{V_R}
\end{equation}
Note that the eigenvalue $\omega$ also satisfies Eq.~(\ref{omega_sym}).

Although the radial potential $V_R$ is singular at the throat $r=C/2$, 
it was shown in \cite{Gonzalez:2008wd} that Eq.~\ref{Z_radial} 
can be transformed into a regular equation. 
However, for the purpose of calculating the spectra, 
we will just integrate the system of equations (\ref{eq_rad}) 
in order to obtain the radial unstable modes.

\section{Numerical Methods}

\subsection{Direct Integration}

For the study of the quasinormal modes describing the ringdown phase, we are interested in gravitational waves being radiated away from the wormhole. From the form of the  Eqs.~(\ref{Z_scalar}), (\ref{Z_axial}) and (\ref{Z_radial}), we can see that asymptotically as $r\to \infty$, the outgoing perturbation will behave like
\begin{eqnarray}
r\to \infty \Rightarrow Z(r,t) \sim e^{i\omega (r_{*} - t)} \sim e^{i\omega (r - t)}  \,.
\end{eqnarray}
On the other side of the throat of the wormhole, we also want the wave to be outgoing at infinity, meaning
\begin{eqnarray}
r\to -\infty \Rightarrow Z(r,t) \sim e^{i\omega (-r_{*} - t)} \sim e^{i\omega (-\zeta r - t)} \,,
\end{eqnarray}
where $\zeta = e^{C\pi/r_0}$ is introduced because of the coordinates not being asymptotically flat at this side of the throat.

In practice these conditions mean that an unstable perturbation 
will have $\omega_I<0$, i.e., the perturbation grows exponentially with time, 
and the perturbation function $Z(r)$ vanishes for $r \to \pm \infty$.
To obtain such unstable modes for the radial perturbations 
we integrate the second--order ODE together with the auxiliary differential equation, 
\begin{equation}
 \frac{d E}{d r} = 0 \,,
\end{equation}
where we promote $\omega$ to a trivial auxiliary function $E \equiv \omega^2$.
The system is supplemented with the following boundary conditions,
\begin{equation}
Z(-\infty) = Z(+ \infty) = 0 \,, \quad \, Z(r_c)=1\,,
\end{equation}
where $r_c$ is an arbitrary point $-\infty<r_c<\infty$.

On the other hand, stable perturbations are damped exponentially as time passes, meaning $\omega_I > 0$, and the radial perturbation function $Z(r)$ explodes exponentially at both infinities. 

In order to obtain the quasinormal modes of the axial and scalar perturbations
numerically, we divide the space at some value $r_c$ into two distinct parts.
For $r>r_c$ , we parametrize the asymptotic behaviour of the perturbation function $Z(r)$ for $r \to +\infty$ according to \cite{Chandrasekhar:1975zza}
\begin{align}
r>r_c \,, \quad  Z^+(r) &=  e^{i \omega r_{*}} Z_P(r)  \,.
\end{align}
Here the leading terms in the asymptotic expansion of $Z_P(r)$ are 
\begin{eqnarray}
&& Z_P (r) = a_0\left(1 + i \, \frac{ l (l+1)}{2 \omega}\frac{1}{r} \right. \\ && \left. + \frac{1}{4 \omega} \left[ -i C (l^2+l+K) - \frac{l (l+2) (l^2-1)}{2 \omega}  \right]\frac{1}{r^2} + ... \right) \,,  \nonumber
\end{eqnarray}
where $a_0$ is the arbitrary amplitude and $K=-1,3$ for scalar and axial perturbations, respectively.

For $r<r_c$, we parametrize the asymptotic behaviour for $r \to -\infty$
according to
\begin{align}
r<r_c \,, \quad  Z^-(r) &=  e^{-i \omega r_{*}} Z_N(r)  \,.
\end{align}
Now the leading terms in the asymptotic expansion of $Z_N(r)$ are 
\begin{eqnarray}
&& Z_N (r) = a_0\left(1 - i \, \frac{ l (l+1)}{2 \omega e^{C \pi/r_0}}\frac{1}{r} \right. \\ && \left. + \frac{1}{4 \omega e^{C \pi/r_0}} \left[ i C (l^2+l+K) - \frac{l (l+2) (l^2-1)}{2 \omega e^{C \pi/r_0}}  \right]\frac{1}{r^2} + ... \right) \,.  \nonumber
\end{eqnarray}

For a particular value of $\omega$, we generate solutions of the functions $Z_N(r)$ and $Z_P(r)$, satisfying the expansions close to the infinities, and $Z_P(r_c)=Z_N(r_c)=1$. The quasinormal modes are obtained when the condition
\begin{equation}
 \frac{1}{Z^-} \frac{d Z^-}{d r_{*}}  \bigg|_{r=r_c} - \frac{1}{Z^+} \frac{d Z^+}{d r_{*}}  \bigg|_{r=r_c}  =0 \,,
\end{equation}
is satisfied.

In order to integrate numerically the equations imposing the corresponding boundary conditions, we use the package Colsys \cite{colsys}.

\subsection{WKB Method}

In addition to the previous numerical method by direct integration of the perturbation equations, we have studied
the spectrum of the axial and scalar perturbations using the WKB method up to third order \cite{Iyer:1986np,Simone:1991wn}. 
In this case the eigenvalue $\omega$ can be approximated by the expression
\begin{equation}
\omega^2 = V_0 + \sqrt{-2 V_0^{(2)}} \Lambda - i \left( n+ \frac{1}{2}  \right) \sqrt{-2 V_0^{(2)}}  (1+\Omega) \,,
\end{equation}
where
\begin{eqnarray}
&& \Lambda = \tfrac{1}{8\sqrt{-2 V_0^{(2)}}} \left[ \left( \tfrac{V_0^{(4)}}{V_0^{(2)}} \right) \left( \tfrac{1}{4}+ \alpha^2 \right)-\tfrac{1}{36}  \left( \tfrac{V_0^{(3)}}{V_0^{(2)}} \right)^2 (7+60 \alpha^2)  \right]   \,,  
\nonumber \\
&& \Omega = \tfrac{1}{-2 V_0^{(2)}} \left[ 
- \tfrac{1}{384} \left( \tfrac{(V_0^{(3)})^{2} V_0^{(4)} }{(V_0^{(2)})^{3}} \right)  (51+100 \alpha^2) 
 \right.  \nonumber 
\\ 
&& \left. 
+ \tfrac{1}{2304} \left( \tfrac{V_0^{(4)}}{V_0^{(2)}} \right)^2  (67+68 \alpha^2)
 \right.  + 
\tfrac{5}{6912}  \left( \tfrac{V_0^{(3)}}{V_0^{(2)}} \right)^4 (77+188 \alpha^2)
\nonumber 
\\ &&
 + \tfrac{1}{284}  \left( \tfrac{V_0^{(3)} V_0^{(5)} }{(V_0^{(2)})^{2}} \right)  (19+28 \alpha^2)  \left. - \tfrac{1}{288}  \left( \tfrac{V_0^{(6)} }{V_0^{(2)}} \right)^2  (5+4 \alpha^2)     \right] \,, 
\end{eqnarray}
with the $(J)$-th derivative of the potential evaluated at the maximum of $V$
\begin{align}
 V_0^{(J)} = \frac{d^{J} V}{d r_{*}^{J}} \bigg|_{r_{*}=r_{*} (r_{\text{max}}) } \,,
\end{align}
and the excitation number $n$ 
\begin{align}
&& \alpha &= n + \frac{1}{2}\,,  \ \ \  n=
\begin{cases}
    0,1,2,...       &  \text{if } \omega_R > 0  \,, \\
    -1, -2, ...  &  \text{if }  \omega_R < 0  \,.
  \end{cases}  
\end{align}
Here we will focus on the fundamental modes, with $n=0$.

Let us note that the WKB method breaks down in a particular case of the axial potential. For $C=0$ (massless wormhole), the peak of the potential is at $r=0$, and the second derivative is
\begin{equation}
 \frac{d^2 V}{d r_{*}^2}\bigg|_{r=0} = \frac{2}{r_0^4} ( 6- l (l+1) ) \,,
\end{equation}
which vanishes when $l=2$. Hence the WKB approximation of the $l=2$ mode 
diverges when the wormhole mass vanishes.

\section{Results and Discussion}

With the use of the methods described above, 
we have calculated the spectrum of modes for each type of perturbation. 
Let us now discuss our results for all three types,
starting with the radial perturbations, 
that give rise to the well-known radial instability
\cite{Shinkai:2002gv,Gonzalez:2008wd,Gonzalez:2008xk}.

\subsection{Unstable radial mode}

\begin{figure}
\centering
\includegraphics[angle =-90,width=0.48\textwidth]{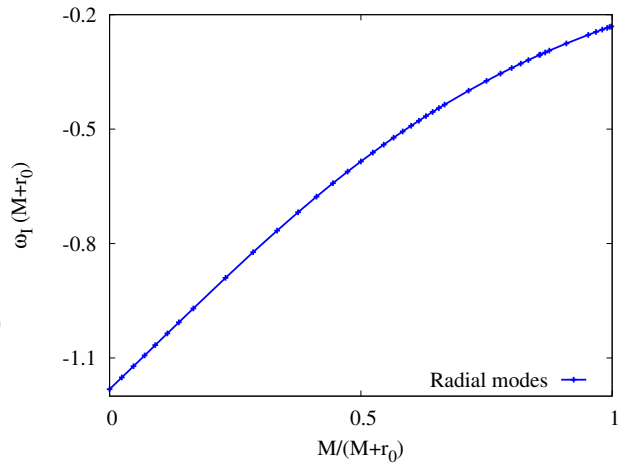}
\caption{Scaled eigenvalue $\omega_I (M+r_0)$ vs the scaled mass $M/(M+r_0)$ 
for the unstable radial mode of the wormhole ($r_0=1$).}
\label{Fig:radial_mode}
\end{figure}

In Fig.~\ref{Fig:radial_mode} we show the scaled imaginary part 
for the eigenvalue $\omega_I (M+r_0)$ vs the scaled mass $M/(M+r_0)$. 
The unstable mode extends from the case of a massless wormhole at $M=0$ 
to the case of an infinitely massive one at $M/(M+r_0)=1$. 
We have fixed $r_0=1$ in the figure, 
although this parameter can always be rescaled 
by a change of the radial coordinate and mass. 
We here focus on the results for positive values of $C$, since
negative values are obtained from the positive ones via
Eq.~(\ref{omega_sym}). 

The unstable mode possesses a purely 
imaginary eigenvalue, i.e., $\omega_R=0$, 
thus the mode increases exponentially as time passes. 
The figure shows that the smallest value of $\omega_I$ 
is obtained for the massless wormhole, where $\omega_I = -1.182$. 
On the other hand, $\omega_I$ approaches zero with increasing mass, 
decaying as $\omega_I \sim 1/M$. 
In fact, extrapolating the numerical data 
yields $\lim_{M\to\infty} \omega_I M \approx -0.23003$.

\subsection{Scalar Perturbations}

Next we present the spectrum of quasinormal modes for scalar perturbations. The modes in this case always correspond to stable perturbations that are exponentially damped in time. 
 
\begin{figure}
\centering
\includegraphics[angle =-90,width=0.48\textwidth]{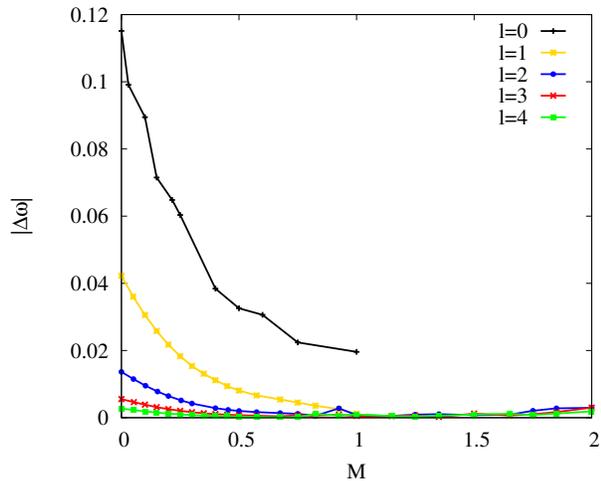}
\caption{Difference between the scalar modes obtained by direct integration 
and the WKB method vs the wormhole mass $M$. Multipole numbers 
$l=0,1,2,3$ and $4$ are shown in black, yellow, blue, red and green, 
respectively.}
\label{Fig:scalar_comp}
\end{figure}

In Fig.~\ref{Fig:scalar_comp} we show the difference between the modes calculated with the direct integration method and the WKB, as a function of the mass of the wormhole $M$. In general what we can see is that the larger the $l$ number, the closer is the WKB method to the direct integration. For $l=0$ the difference is the largest. In general increasing the mass also reduces the difference between the two methods, 
making the WKB a better approximation for the more massive wormholes. 
Note, however, that in general the direct integration method loses precision 
for mass $M\gtrsim 1.5$. 
In fact, for $l=0,1$ the direct integration method becomes rather unstable 
when $M\gtrsim 1$, 
and we do not include the results in the figure. 
Thus for larger masses the WKB seems to become more precise 
than the direct integration method.
Overall the qualitative features of the spectrum are very well approximated by the WKB method in all cases. 

\begin{figure}
\centering
\includegraphics[angle =-90,width=0.48\textwidth]{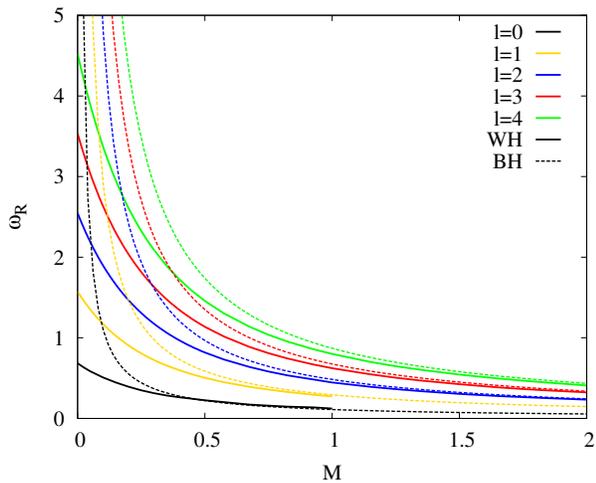}
\caption{Real part of the eigenvalue,
$\omega_R$, vs the mass $M$ for the scalar modes ($r_0=1$),
obtained by direct integration.
Multipole numbers
$l=0,1,2,3$ and $4$ are shown in black, yellow, blue, red and green,
respectively.
Solid lines correspond to the wormhole modes,
dashed lines to the Schwarzschild black hole modes.  
}
\label{Fig:scalar_omegaR}
\end{figure}

In Fig.~\ref{Fig:scalar_omegaR} we show the real part of the eigenfrequency as a function of the mass. We focus on positive values. Solid lines correspond to the modes of the wormhole, and dashed lines correspond to the modes of the Schwarzschild black hole of the same mass. In the figure we also include results for different multipole numbers: $l=0,1,2,3$ and $4$ in black, yellow, blue, red and green, respectively. Note that the modes of the Schwarzschild black hole behave like $\omega M = const$. Hence they diverge at $M=0$. 

The value of $\omega_R$ for a wormhole of a given mass $M$ 
is always below the corresponding value of a black hole of the same mass, except for $l=0$, where they cross at $M=0.46$ . 
The difference between wormholes and black holes decreases 
as the mass increases. 
Note that for a fixed value of the mass, the difference increases 
as the multipole number $l$ increases.

\begin{figure}
\centering
\includegraphics[angle =-90,width=0.48\textwidth]{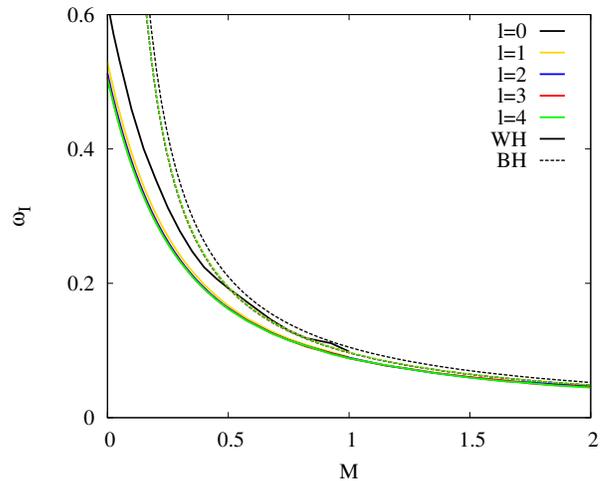}
\caption{Analogous to Fig.~\ref{Fig:scalar_omegaR} for 
the imaginary part of the eigenvalue, $\omega_I$.  }
\label{Fig:scalar_omegaI}
\end{figure}

In Fig.~\ref{Fig:scalar_omegaI} we show the analogous figure 
for the imaginary part of the eigenvalue $\omega$. 
For a fixed value of the mass $M$, the value of $\omega_I$ 
for the wormhole is always smaller than the value for the black hole. 
It is interesting to note that the imaginary part does not depend 
very much on the multipole number $l$. 
Hence the modes of the wormhole are very close to one another, 
with the largest difference for the multipole $l=0$. 
Moreover, as the mass is increased, 
the wormhole modes come closer and closer
to the corresponding Schwarzschild modes.

From these figures we conclude that very large wormholes possess 
a quasinormal mode spectrum very similar to the one of black holes. 
Hence the ringdown phase of a purely scalar perturbation will look very similar to the ringdown of a test scalar in the background of a Schwarzschild black hole.

However, the differences between both spectra become larger and larger 
as the mass is reduced. Thus small wormholes possess a ringdown 
with smaller frequencies ($\omega_R$)  
and larger damping times ($1/\omega_I$) 
than a test field in the background of a black hole.

\subsection{Axial Perturbations}

We now turn to the spectrum of quasinormal modes of axial perturbations. 
As in the previous case of scalar perturbations, the modes always correspond to stable perturbations that are exponentially damped in time. 

\begin{figure}
\centering
\includegraphics[angle =-90,width=0.48\textwidth]{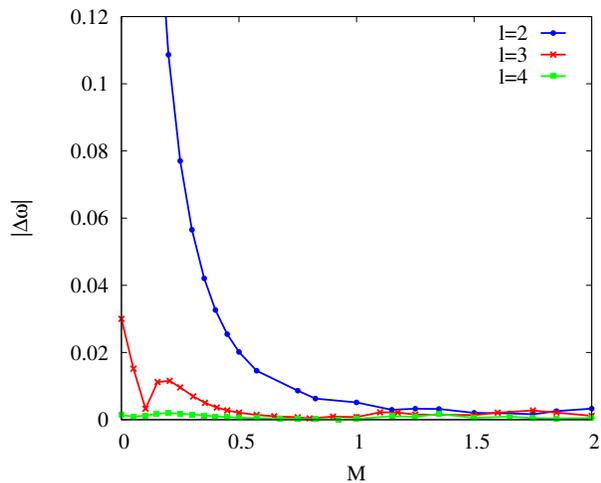}
\caption{
Difference between the axial modes obtained by direct integration
and the WKB method vs the wormhole mass $M$. Multipole numbers
$l=2,3$ and $4$ are shown in blue, red and green,
respectively.}
\label{Fig:axial_comp}
\end{figure}

In Fig.~\ref{Fig:axial_comp} we show the difference 
between the modes calculated by direct integration and the WKB method. 
Analogous to Fig.~\ref{Fig:scalar_comp}, 
the larger the multipole number $l$ the closer are both results. 
This remains true as the mass is increased. 
Note, however, the problem of the WKB method in the $l=2$ massless case 
\cite{Kim:2008zzj}. As discussed above, 
in this case the frequency and damping time become singular, 
and the WKB method fails.
 
In fact the calculation of the mode for small masses deviates too much, and can only be trusted for $M\gtrsim 0.1$. Except for these particular cases, the qualitative features of the spectrum are again very well approximated by the WKB method.

\begin{figure}
\centering
\includegraphics[angle =0,width=0.48\textwidth]{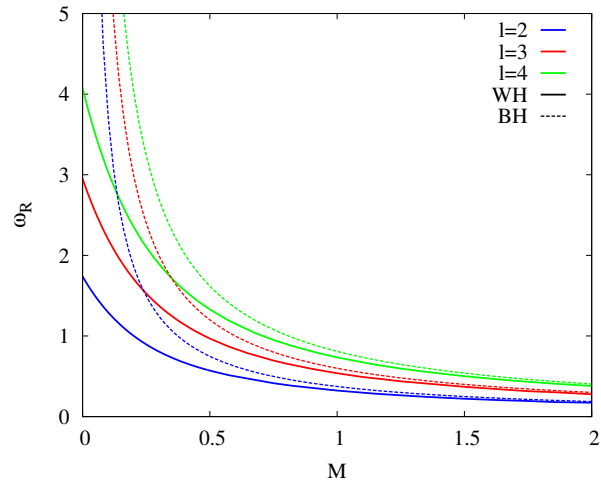}
\caption{
Real part of the eigenvalue,
$\omega_R$, vs the mass $M$ for the axial modes ($r_0=1$),
obtained by direct integration.
Multipole numbers
$l=2,3$ and $4$ are shown in blue, red and green,
respectively.
Solid lines correspond to the wormhole modes,
dashed lines to the Schwarzschild black hole modes.}
\label{Fig:axial_omegaR}
\end{figure}

In Fig.~\ref{Fig:axial_omegaR} we show the real part of the eigenfrequency as a function of the mass, focusing on positive values. We compare the spectrum of the wormhole (with solid lines) with the spectrum of the Schwarzschild black hole (with dashed lines), and colors correspond to different multipole numbers, with $l=2$ in blue, $l=3$ in red and $l=4$ in green. 

The result is very similar to the case of purely scalar perturbations: 
for a fixed value of $M$, the real part of $\omega$ is always below 
the corresponding value for the black hole. 
The difference is largest for the massless case 
(for which the frequency of the black hole diverges, 
while it is finite for the wormhole). 
However, as the mass is increased, the spectra come closer to each other, 
and the modes of the wormhole become very well approximated 
by the modes of the black hole. 
Note, that when the value of the mass is fixed, 
an increase of the multipole number $l$ leads to an increase of
the difference between the spectra. 
Hence wormholes and black holes could be distinguished
by comparing their high multipole spectra.

\begin{figure}
\centering
\includegraphics[angle =0,width=0.48\textwidth]{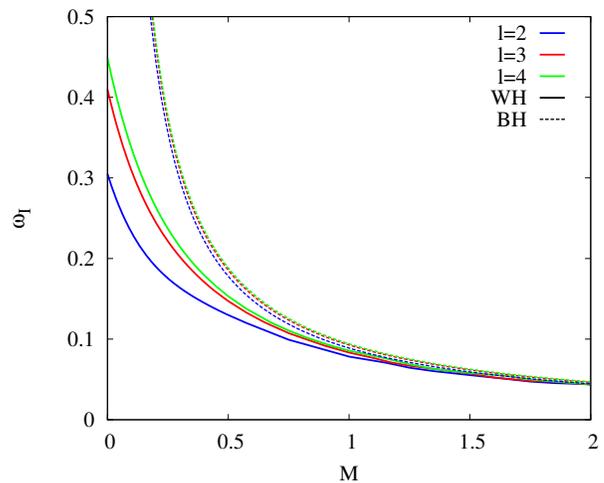}
\caption{
Analogous to Fig.~\ref{Fig:axial_omegaR} for
the imaginary part of the eigenvalue, $\omega_I$.}
\label{Fig:axial_omegaI}
\end{figure}

In Fig.~\ref{Fig:axial_omegaI} we show the imaginary part of $\omega$ 
vs the mass. 
The behaviour is slightly different from the scalar case 
(Fig.~\ref{Fig:scalar_omegaI}), 
for which the imaginary part did not depend very much
on the multipole number. 
In the axial case, increasing the multipole number $l$ 
leads to an increase of the value of $\omega_I$ (for fixed $M$). 
Nevertheless, the value of $\omega_I$ of a wormhole 
is also always below the value of $\omega_I$ of a black hole.

Hence the results are qualitatively similar to the scalar case.
The frequencies that will appear during the ringdown phase 
of an axial perturbation will be smaller 
in the case of a wormhole as compared to a black hole of the same mass. 
On the other hand the damping times are larger. 
However, very large wormholes will have spectra 
very similar to very large black holes. 

\subsection{Comparison of the spectra of massive wormholes and massive black holes}

\begin{figure}
\centering
\includegraphics[angle =-90,width=0.48\textwidth]{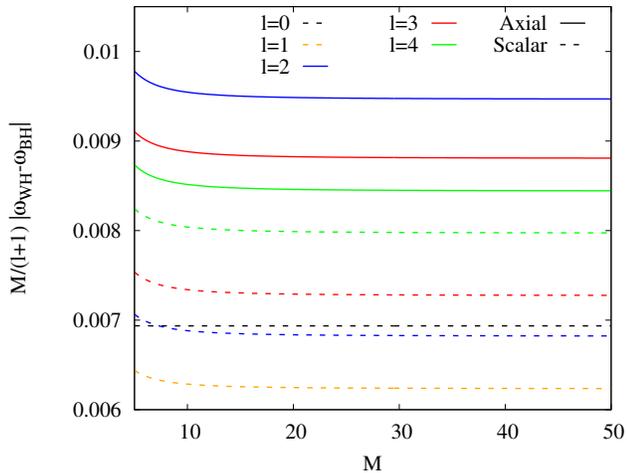}
\caption{Difference between the quasinormal modes of wormholes and
black holes for large values of the mass using the WKB method.}
\label{Fig:largeM}
\end{figure}

In the previous subsections we have seen 
that the spectra of massive wormholes come closer and closer 
to the spectra of Schwarzschild black holes as the mass is increased. 
Although the accuracy of the employed method of direct integration 
does not allow us to calculate the modes with sufficiently high precision 
in the large $M$ limit, in a first approximation 
we can compare the analytical results 
for wormholes and black holes, obtained in the eikonal approximation.
Note, that the effective potentials coincide for large $l$
for scalar and axial perturbations.

In the eikonal approximation ($l \gg 1$), 
a massive black hole ($M \gg 1$) possesses the following approximate spectrum
\begin{equation}
 \omega_{BH} = \frac{1}{3 \sqrt{3} M} (l+1) - i \left( n+\frac{1}{2}\right) \frac{1}{3 \sqrt{3} M} \,.
\end{equation}
In the case of a wormhole, using the WKB formulas 
(or equivalently the geodesic approach) 
and considering $l \gg 1$ leads to the approximate eikonal spectrum
\begin{equation}
 \omega_{WH} = \frac{1}{2 e M} (l+1) - i \left( n+\frac{1}{2} \right) \frac{1}{2 e M} \, ,
\end{equation}
where $e$ is Euler's number. Hence one finds for the fundamental mode
\begin{eqnarray}
 \Re(\omega_{BH})-\Re(\omega_{WH}) = (\frac{1}{3 \sqrt{3}}- \frac{1}{2 e})\frac{l+1}{M}, \nonumber \\
 \Im(\omega_{BH})-\Im(\omega_{WH}) = -(\frac{1}{6 \sqrt{3}}- \frac{1}{4 e})\frac{1}{M}.
\end{eqnarray}
Thus for large $l$ the modes approach each other as $M$ increases, 
which is compatible with what we have obtained 
for the lowest multipole numbers, as well. 

It is interesting to note that by considering the scaled modes 
$\bar \omega_R = \omega_R M/(l+1)$ and $\bar \omega_I = \omega_I M$
in the eikonal approximation, 
the difference 
between the wormholes and black holes
for each of these quantities is only $0.85\%$ and $0.43\%$, respectively.

In the case of small values of $l$, the WKB approximation 
can be used directly to make such a comparison. 
Recall that in the large $M$ regime
this approximation is most likely better 
than the employed direct integration method.
In Fig.~\ref{Fig:largeM} we show the difference 
between the wormhole modes and the black hole modes vs the mass,
obtained in the WKB approximation.
The result is similar to what the eikonal approximation shows: 
the difference approaches constant values, 
and the order of magnitude is very similar 
to what the eikonal approximation predicts, 
with a difference of less than $1\%$. 
It is interesting to note that the difference in the scalar modes is smaller than the difference in the axial modes.

\section{Conclusions}

We have studied radial, scalar and axial modes of massive static 
wormholes supported by a massless phantom field.
The background solutions depend on two parameters, 
a throat parameter $r_0$ and a symmetry parameter $C$,
proportional to the mass $M$ of the wormhole.
Since the eigenvalues $\omega(C)$ and $\omega(-C)$ satisfy
a simple relation, Eq.~(\ref{omega_sym}),
we have focussed on the wormholes with mass $M \ge 0$.
To obtain the modes
we have employed direct numerical integration as well as the
WKB method and compared the results from both methods.

For the radial perturbations we have evaluated the eigenvalue 
$\omega$ 
of the unstable radial mode, which is a purely imaginary mode
with a negative imaginary part $\omega_I$,
i.e., the mode increases exponentially in time.
The radial instability is well-known for vanishing wormhole mass
\cite{Shinkai:2002gv,Gonzalez:2008wd,Gonzalez:2008xk}.
Here we have determined its dependence on the wormhole mass.

We have obtained the spectrum of the scalar perturbations 
for multipole numbers $l=0, \dots, 4$ and small masses,
employing and comparing both methods. In fact, the larger 
the multipole number $l$,
the smaller the difference between the methods becomes.
Comparing the scalar wormhole modes with the corresponding
scalar Schwarzschild black hole modes 
we note, that their difference decreases with increasing mass.
Thus for large masses both spectra become very similar.
In contrast, for small masses their difference grows strongly,
since the black hole modes diverge for vanishing mass,
while the wormhole modes stay finite.

We have analyzed the axial perturbations
for multipole numbers $l=2,3$ and 4.
Like the scalar perturbations, the axial perturbations
always lead to stable modes.
Whereas the $l=2$ case is special, since the WKB
approximation breaks down for vanishing mass,
for higher multipole numbers $l$ the difference between
both methods is again overall small and decreases with increasing $l$.
Comparison of the axial wormhole and black hole modes
leads to similar conclusions as in the scalar case.

Since the direct integration method employed
seems to become less reliable for large masses,
while being rather close to the WKB method before,
we have addressed the high mass limit of the
scalar and axial modes by considering the eikonal
approximation for large multipole numbers
and ordinary WKB for small multipole numbers.
We have found, that for fixed multipole number $l$
the difference between the modes decreases with increasing mass
as $1/M$, but there are differences in the scaled spectra. 
In particular, the difference between
the eikonal approximation and the WKB can be about $1\%$.

Concerning future observations of gravitational waves,
it appears that the best chances to distiguish wormholes
and black holes would be to consider
low masses and low multipole numbers $l$, 
since the spectra become very different in that case.

\section*{Acknowledgements}

XYC would like to thank Kevin Eickhoff for useful discussions.
The authors would like to acknowledge support from the DFG Research Training Group 1620 {\sl Models of Gravity} 
and the
JLBS and JK would like to acknowledge support by the COST Action CA16104.
JLBS would like to acknowledge support from the DFG project BL 1553.


\begin{thebibliography}{99}

\bibitem{Morris:1988cz}
  M.~S.~Morris, K.~S.~Thorne,
  Am.\ J.\ Phys.\  {\bf 56}, 395 (1988).

\bibitem{Visser:1995cc}
  M.~Visser,
  ``Lorentzian wormholes: From Einstein to Hawking,''
  (AIP, Woodbury, USA 1995)

\bibitem{Lobo:2017eum}
  F.~S.~N.~Lobo,
  ``Wormholes, Warp Drives and Energy Conditions'',
  Springer series  Fundam.\ Theor.\ Phys.\  {\bf 189} (2017).



\bibitem{Einstein:1935tc}
  A.~Einstein and N.~Rosen,
  Phys.\ Rev.\  {\bf 48} (1935) 73.

\bibitem{Lobo:2005us}
  F.~S.~N.~Lobo,
  Phys.\ Rev.\ D {\bf 71}, 084011 (2005).

\bibitem{Ellis:1973yv}
  H.~G.~Ellis,
  J.\ Math.\ Phys.\  {\bf 14}, 104 (1973).

\bibitem{Bronnikov:1973fh}
  K.~A.~Bronnikov,
  Acta Phys.\ Polon.\ B {\bf 4}, 251 (1973).

\bibitem{Ellis:1979bh}
  H.~G.~Ellis,
  Gen.\ Rel.\ Grav.\  {\bf 10}, 105 (1979).

\bibitem{Torii:2013xba} 
  T.~Torii and H.~a.~Shinkai,
  Phys.\ Rev.\ D {\bf 88}, 064027 (2013)

\bibitem{Dzhunushaliev:2013jja} 
  V.~Dzhunushaliev, V.~Folomeev, B.~Kleihaus, J.~Kunz and E.~Radu,
  Phys.\ Rev.\ D {\bf 88}, 124028 (2013)

\bibitem{Kashargin:2007mm}
  P.~E.~Kashargin and S.~V.~Sushkov,
  Grav.\ Cosmol.\  {\bf 14}, 80 (2008).

\bibitem{Kashargin:2008pk}
  P.~E.~Kashargin and S.~V.~Sushkov,
  Phys.\ Rev.\ D {\bf 78}, 064071 (2008).

\bibitem{Kleihaus:2014dla}
  B.~Kleihaus and J.~Kunz,
  Phys.\ Rev.\ D {\bf 90} (2014) 121503

\bibitem{Chew:2016epf}
  X.~Y.~Chew, B.~Kleihaus and J.~Kunz,
  Phys.\ Rev.\ D {\bf 94}, no. 10, 104031 (2016)

\bibitem{Chew:2018vjp}
  X.~Y.~Chew, B.~Kleihaus and J.~Kunz,
  Phys.\ Rev.\ D {\bf 97} (2018) 064026

\bibitem{Dzhunushaliev:2017syc}
  V.~Dzhunushaliev, V.~Folomeev, B.~Kleihaus and J.~Kunz,
  Phys.\ Rev.\ D {\bf 97} (2018) no.2,  024002





\bibitem{Cramer:1994qj}
  J.~G.~Cramer, R.~L.~Forward, M.~S.~Morris, M.~Visser, G.~Benford and G.~A.~Landis,
  Phys.\ Rev.\ D {\bf 51} (1995) 3117

\bibitem{Safonova:2001vz}
  M.~Safonova, D.~F.~Torres and G.~E.~Romero,
  Phys.\ Rev.\ D {\bf 65} (2002) 023001

\bibitem{Perlick:2003vg}
  V.~Perlick,
  Phys.\ Rev.\ D {\bf 69} (2004) 064017

\bibitem{Nandi:2006ds}
  K.~K.~Nandi, Y.~Z.~Zhang and A.~V.~Zakharov,
  Phys.\ Rev.\ D {\bf 74} (2006) 024020

\bibitem{Abe:2010ap}
  F.~Abe,
  Astrophys.\ J.\  {\bf 725}, 787 (2010).

\bibitem{Toki:2011zu}
  Y.~Toki, T.~Kitamura, H.~Asada and F.~Abe,
  Astrophys.\ J.\  {\bf 740} (2011) 121

\bibitem{Nakajima:2012pu}
  K.~Nakajima and H.~Asada,
  Phys.\ Rev.\ D {\bf 85} (2012) 107501

\bibitem{Tsukamoto:2012xs}
  N.~Tsukamoto, T.~Harada and K.~Yajima,
  Phys.\ Rev.\ D {\bf 86} (2012) 104062

\bibitem{Kuhfittig:2013hva}
  P.~K.~F.~Kuhfittig,
  Eur.\ Phys.\ J.\ C {\bf 74} (2014) no.99,  2818

\bibitem{Bambi:2013nla}
  C.~Bambi,
  Phys.\ Rev.\ D {\bf 87} (2013) 107501

\bibitem{Takahashi:2013jqa}
  R.~Takahashi and H.~Asada,
  Astrophys.\ J.\  {\bf 768} (2013) L16

\bibitem{Tsukamoto:2016zdu}
  N.~Tsukamoto and T.~Harada,
  Phys.\ Rev.\ D {\bf 95} (2017) no.2,  024030

\bibitem{Nedkova:2013msa}
  P.~G.~Nedkova, V.~K.~Tinchev and S.~S.~Yazadjiev,
  Phys.\ Rev.\ D {\bf 88} (2013) no.12,  124019

\bibitem{Ohgami:2015nra} 
  T.~Ohgami and N.~Sakai,
  Phys.\ Rev.\ D {\bf 91}, 124020 (2015).

\bibitem{Shaikh:2018kfv} 
  R.~Shaikh,
  arXiv:1803.11422 [gr-qc].

\bibitem{Gyulchev:2018fmd}
  G.~Gyulchev, P.~Nedkova, V.~Tinchev and S.~Yazadjiev,
  arXiv:1805.11591 [gr-qc].

\bibitem{Harko:2008vy} 
  T.~Harko, Z.~Kovacs and F.~S.~N.~Lobo,
  Phys.\ Rev.\ D {\bf 78}, 084005 (2008)

\bibitem{Harko:2009xf} 
  T.~Harko, Z.~Kovacs and F.~S.~N.~Lobo,
  Phys.\ Rev.\ D {\bf 79}, 064001 (2009)

\bibitem{Bambi:2013jda} 
  C.~Bambi,
  Phys.\ Rev.\ D {\bf 87}, 084039 (2013)

\bibitem{Zhou:2016koy}
  M.~Zhou, A.~Cardenas-Avendano, C.~Bambi, B.~Kleihaus and J.~Kunz,
  Phys.\ Rev.\ D {\bf 94}, 024036 (2016).

\bibitem{Lamy:2018zvj}
  F.~Lamy, E.~Gourgoulhon, T.~Paumard and F.~H.~Vincent,
  arXiv:1802.01635 [gr-qc].



\bibitem{Damour:2007ap} 
  T.~Damour and S.~N.~Solodukhin,
  Phys.\ Rev.\ D {\bf 76}, 024016 (2007).

\bibitem{Azreg-Ainou:2014dwa} 
  M.~Azreg-Aïnou,
  JCAP {\bf 1507}, 037 (2015)

\bibitem{Dzhunushaliev:2016ylj} 
  V.~Dzhunushaliev, V.~Folomeev, B.~Kleihaus and J.~Kunz,
  JCAP {\bf 1608}, no. 08, 030 (2016)

\bibitem{Cardoso:2016rao} 
  V.~Cardoso, E.~Franzin and P.~Pani,
  Phys.\ Rev.\ Lett.\  {\bf 116}, 171101 (2016)
  Erratum: [Phys.\ Rev.\ Lett.\  {\bf 117}, 089902 (2016)].

\bibitem{Konoplya:2016hmd} 
  R.~A.~Konoplya and A.~Zhidenko,
  JCAP {\bf 1612}, 043 (2016).

\bibitem{Nandi:2016uzg} 
  K.~K.~Nandi, R.~N.~Izmailov, A.~A.~Yanbekov and A.~A.~Shayakhmetov,
  Phys.\ Rev.\ D {\bf 95}, 104011 (2017).

\bibitem{Bueno:2017hyj} 
  P.~Bueno, P.~A.~Cano, F.~Goelen, T.~Hertog and B.~Vercnocke,
  Phys.\ Rev.\ D {\bf 97}, 024040 (2018)

\bibitem{Dzhunushaliev:2011xx}
  V.~Dzhunushaliev, V.~Folomeev, B.~Kleihaus and J.~Kunz,
  JCAP {\bf 1104}, 031 (2011).

\bibitem{Dzhunushaliev:2012ke}
  V.~Dzhunushaliev, V.~Folomeev, B.~Kleihaus and J.~Kunz,
  Phys.\ Rev.\ D {\bf 85}, 124028 (2012).

\bibitem{Dzhunushaliev:2013lna}
  V.~Dzhunushaliev, V.~Folomeev, B.~Kleihaus and J.~Kunz,
  Phys.\ Rev.\ D {\bf 87}, 104036 (2013).

\bibitem{Dzhunushaliev:2014mza}
  V.~Dzhunushaliev, V.~Folomeev, B.~Kleihaus and J.~Kunz,
  Phys.\ Rev.\ D {\bf 89}, 084018 (2014)

\bibitem{Aringazin:2014rva}
  A.~Aringazin, V.~Dzhunushaliev, V.~Folomeev, B.~Kleihaus and J.~Kunz,
  JCAP {\bf 1504} (2015) no.04,  005

\bibitem{Dzhunushaliev:2014bya}
  V.~Dzhunushaliev, V.~Folomeev, C.~Hoffmann, B.~Kleihaus and J.~Kunz,
  Phys.\ Rev.\ D {\bf 90} (2014) no.12,  124038

\bibitem{Hoffmann:2017jfs}
  C.~Hoffmann, T.~Ioannidou, S.~Kahlen, B.~Kleihaus and J.~Kunz,
  Phys.\ Rev.\ D {\bf 95} (2017) no.8,  084010

\bibitem{Hoffmann:2017vkf}
  C.~Hoffmann, T.~Ioannidou, S.~Kahlen, B.~Kleihaus and J.~Kunz,
  Phys.\ Lett.\ B {\bf 778} (2018) 161

\bibitem{Hoffmann:2018}
  C.~Hoffmann, T.~Ioannidou, S.~Kahlen, B.~Kleihaus and J.~Kunz,
  arXiv:1803.11044 [gr-qc].




\bibitem{Abbott:2016blz}
  B.~P.~Abbott {\it et al.} [LIGO Scientific and Virgo Collaborations],
  Phys.\ Rev.\ Lett.\  {\bf 116} (2016) 061102

\bibitem{Abbott:2016nmj}
  B.~P.~Abbott {\it et al.} [LIGO Scientific and Virgo Collaborations],
  Phys.\ Rev.\ Lett.\  {\bf 116} (2016) 241103

\bibitem{Abbott:2017vtc}
  B.~P.~Abbott {\it et al.} [LIGO Scientific and VIRGO Collaborations],
  Phys.\ Rev.\ Lett.\  {\bf 118} (2017) 221101

\bibitem{Abbott:2017oio} 
  B.~P.~Abbott {\it et al.} [LIGO Scientific and Virgo Collaborations],
  Phys.\ Rev.\ Lett.\  {\bf 119}, 141101 (2017)

\bibitem{TheLIGOScientific:2017qsa} 
  B.~P.~Abbott {\it et al.} [LIGO Scientific and Virgo Collaborations],
  Phys.\ Rev.\ Lett.\  {\bf 119}, 161101 (2017)

\bibitem{Kim:2008zzj}
  S.~W.~Kim,
  Prog.\ Theor.\ Phys.\ Suppl.\  {\bf 172}, 21 (2008).

\bibitem{Konoplya:2005et} 
  R.~A.~Konoplya and C.~Molina,
  Phys.\ Rev.\ D {\bf 71}, 124009 (2005)

\bibitem{Konoplya:2010kv} 
  R.~A.~Konoplya and A.~Zhidenko,
  Phys.\ Rev.\ D {\bf 81}, 124036 (2010)

\bibitem{Aneesh:2018hlp}
  S.~Aneesh, S.~Bose and S.~Kar,
  arXiv:1803.10204 [gr-qc].

\bibitem{Volkel:2018hwb} 
  S.~H.~V\"olkel and K.~D.~Kokkotas,
  Class.\ Quant.\ Grav.\  {\bf 35}, no. 10, 105018 (2018)

\bibitem{Shinkai:2002gv}
  H.~a.~Shinkai and S.~A.~Hayward,
  Phys.\ Rev.\ D {\bf 66} (2002) 044005

\bibitem{Gonzalez:2008wd}
  J.~A.~Gonzalez, F.~S.~Guzman and O.~Sarbach,
  Class.\ Quant.\ Grav.\  {\bf 26} (2009) 015010

\bibitem{Gonzalez:2008xk}
  J.~A.~Gonzalez, F.~S.~Guzman and O.~Sarbach,
  Class.\ Quant.\ Grav.\  {\bf 26} (2009) 015011

\bibitem{Cremona:2018wkj} 
  F.~Cremona, F.~Pirotta and L.~Pizzocchero,
  arXiv:1805.02602 [gr-qc].

\bibitem{Matos:2005uh}
  T.~Matos and D.~Nunez,
  Class.\ Quant.\ Grav.\  {\bf 23}, 4485 (2006).



\bibitem{Cardoso:2001bb} 
  V.~Cardoso and J.~P.~S.~Lemos,
  Phys.\ Rev.\ D {\bf 64}, 084017 (2001)

\bibitem{Chandrasekhar:1975zza}
  S.~Chandrasekhar and S.~L.~Detweiler,
  Proc.\ Roy.\ Soc.\ Lond.\ A {\bf 344} (1975) 441.

\bibitem{colsys}
U.~Ascher, J.~Christiansen and R.D.~Russell,
Math.\ Comp. \textbf{33}, 659 (1979)  


\bibitem{Iyer:1986np}
  S.~Iyer and C.~M.~Will,
  Phys.\ Rev.\ D {\bf 35} (1987) 3621.

\bibitem{Simone:1991wn}
  L.~E.~Simone and C.~M.~Will,
  Class.\ Quant.\ Grav.\  {\bf 9} (1992) 963.

\end{thebibliography}
\end{document}